\documentclass[11pt]{article}
\usepackage[utf8]{inputenc}

\usepackage[pdftex,dvipsnames]{xcolor}  
\usepackage{amsmath,amssymb}

\usepackage[margin=1in, top=1in, bottom=1in]{geometry}

\usepackage{newpxtext}
\usepackage[varg,bigdelims]{newpxmath}
\usepackage{array}

\usepackage[backend=biber, backref=true, maxbibnames = 10, style = alphabetic]{biblatex}
\usepackage[bookmarks=true, colorlinks=true, linkcolor=blue!50!black, citecolor=orange!50!black, urlcolor=orange!50!black, pdfencoding=unicode]{hyperref}

\usepackage{listings}

\lstdefinelanguage[2026]{CaMPL}{%
  keywords={},%
  otherkeywords={->,=>,::,|,\,},%
  morekeywords={[1]data,codata,protocol,coprotocol,proc,%
    do,on,with,into,as,%
  },%
  morekeywords={[2]Console},%
  morekeywords={[3]Put,Get,TopBot,Neg,Store},%
  morekeywords={[4]hput,hcase,put,get,halt,plug,close,fork,split,neg,store,use,race},%
  sensitive,%
  morecomment=[l]--,%
  morecomment=[n]{\{-}{-\}},%
  morestring=[b]"%
}[keywords,comments,strings]%
\definecolor{dkgreen}{rgb}{0,0.6,0}
\definecolor{gray}{rgb}{0.5,0.5,0.5}
\definecolor{purple}{rgb}{0.8,0,0.3}
\definecolor{orange}{rgb}{1,0.4,0}
\definecolor{lightlightgray}{rgb}{.95,.95,.95}
\definecolor{lightgray}{rgb}{.9,.9,.9}
\definecolor{lightgray2}{rgb}{.85,.85,.85}
\definecolor{darkgray}{rgb}{.4,.4,.4}
\definecolor{darkred}{rgb}{0.6,0,0}

\lstdefinestyle{CaMPLstyle}{%
  language=CaMPL,%
  keywordstyle={[1]\color{darkred}},%
  keywordstyle={[2]\color{blue}},%
  keywordstyle={[3]\color{dkgreen}},%
  keywordstyle={[4]\color{purple}},%
  commentstyle={\itshape\color{orange}},%
  identifierstyle=\color{black},%
  stringstyle=\color{teal},%
  showstringspaces=false,%
  basicstyle=\ttfamily\small,%
  tabsize=2,%
  showtabs=false,%
  tab={\color{gray}\rightarrowfill},%
  numbers=left,%
  numberstyle=\color{gray},%
  breaklines=true,%
  frame=single,%
  framexleftmargin=0pt,%
  rulecolor=\color{lightgray2},%
  rulesepcolor=\color{gray},%
  backgroundcolor=\color{lightgray},%
}%
\usepackage{tikz}
\usepackage[all,cmtip]{xy}

\tikzstyle{strings}=[baseline={([yshift=-.5ex]current bounding box.center)}]

\tikzset{every picture/.append style={scale=.5}, transform shape, strings}

\tikzset{%
symbol/.style={%
draw=none,
every to/.append style={%
edge node={node [sloped, allow upside down, auto=false]{$#1$}}}
}
}

\usetikzlibrary{shapes.geometric}
\usetikzlibrary{patterns}
\usetikzlibrary{fit}
\usetikzlibrary{positioning}
\usetikzlibrary{calc}
\usetikzlibrary{arrows}
\usetikzlibrary{decorations.markings}
\usetikzlibrary{decorations.pathreplacing}
\usetikzlibrary{shapes}

\pgfdeclarelayer{nodelayer}
\pgfdeclarelayer{edgelayer}
\pgfsetlayers{edgelayer,nodelayer,main}

\tikzstyle{none}=[inner sep=-1pt]
\tikzstyle{circle}=[shape=circle,draw]
\tikzstyle{head}=[fill=white, draw=black, shape=circle, scale=2, thick]

\newcommand{\blackman}{
\begin{tikzpicture}[scale=1.5]
	\begin{pgfonlayer}{nodelayer}
		\node [style=head] (0) at (-3, 5) {};
		\node [style=none] (1) at (-3, 4) {};
		\node [style=none] (2) at (-3.25, 3.75) {};
		\node [style=none] (3) at (-2.75, 3.75) {};
		\node [style=none] (4) at (-3.25, 4.45) {};
		\node [style=none] (5) at (-2.75, 4.45) {};
	\end{pgfonlayer}
	\begin{pgfonlayer}{edgelayer}
		\draw[thick] (0) to (1.center);
		\draw[thick] (1.center) to (2.center);
		\draw[thick] (1.center) to (3.center);
		\draw[thick] (4.center) to (5.center);
	\end{pgfonlayer}
\end{tikzpicture} }

\definecolor{darkMagenta}{HTML}{DE3163}
\newcommand{\tc}[1]{\textcolor{darkMagenta}{\textit{#1}}}
\newcommand{\ttc}[1]{\textcolor{darkMagenta}{\texttt{#1}}}

\newcommand{\comment}[1]{\textcolor{blue}{#1}}
\title{Categorical Message Passing Language (CaMPL) \\ for programmers}

\author{
    Daniel Kiyoshi Hashimoto\thanks{Universidade Federal do Rio de Janeiro, Brazil.} \and Alexanna {Little Berg}\thanks{University of Calgary, Canada.} \and Priyaa Varshinee Srinivasan\thanks{Tallinn University of Technology, Estonia. This work was co-funded by the European Union and Estonian Research Council through the Mobilitas 3.0
(MOB3JD1227).}}
    
\date{\vspace{-3em}}

\begin{document}

\maketitle

\begin{abstract} Categorical Message Passing Language (CaMPL) is a functional-style concurrent programming language whose semantics is in category theory, more specifically, linear actegories. Its core programming feature is message passing along typed communication channels between concurrent processes. CaMPL also supports controlled non-determinism via 'races' which allow processes to adapt dynamically while they are running, higher-order processes which pass other processes as messages, and custom channel datatypes called protocols and coprotocols which allow one to define infinite channel types or implement session types. 

The type system of CaMPL arises from a Curry-Howard-Lambek-like correspondence for concurrent programming, established by Cockett and Pastro in their paper titled ``The logic of message passing''.
This type system ensures that a formal CaMPL program, i.e., one which does not allow general recursion, will never become deadlocked or livelocked. In this article, we explore the type system of CaMPL, custom channel types, and controlled non-determinism using code examples after briefly introducing its mathematical underpinnings.
\end{abstract}

\section{Introduction}

Categorical Message Passing Language (CaMPL) is a functional-style concurrent programming language that uses message-passing concurrency.
It implements a type system based on linear logic that was described in Cockett and Pastro’s paper ``The logic of message passing'' \cite{CoP07}. The primary result of their paper was defining and proving a concurrent analogue of the functional programming Curry-Howard-Lambek correspondence or proof-as-programs principle \cite{Howard1980, Lambek1969, Lambek1972}. They designed a two-tiered logic consisting of a sequential part, called the message logic, that interacts with a concurrent part, called the message-passing logic. 
The categorical semantics of the message-passing logic is given by a {\em linear actegory}.
A linear actegory minimally consists of a monoidal category acting covariantly and contravariantly on a linearly distributive category, 
satisfying certain coherences~\cite{CoP07, CS97}. 

The concurrent type system provided by linear actegories ensures that processes can never be connected in a cycle.
That is, at any instance in run-time, the topology of a program is guaranteed to be a finite acyclic graph consisting of processes, which are nodes, and channels, which are edges.
The benefit of this property is that problems caused by cycles, such as deadlocks and livelocks, do not occur \cite{Ly18}.
However, similar to proving termination in the sequential case, livelock freedom is only guaranteed if we disallow general recursion.
Deadlock freedom is still guaranteed.

The first implementation of the CaMPL compiler was written by Kumar as described in \cite{Ku18}.
This version implemented sequential and concurrent type systems.
It added features to define custom concurrent data types called protocols and coprotocols.
The categorical semantics for protocols and coprotocols were given by Yeasin in \cite{Ye12}.
This compiler performed both a type inference and type checking.
The second implementation of the compiler was written by Pon as described in \cite{Po21, Po22}.
This version re-implemented the existing features and added controlled non-determinism in the form of ``races.''
Changes to the categorical semantics to include non-determinism were outlined by Little in \cite{xanna, xanna2}.
The most recent feature added to the compiler was higher-order message passing which allows processes to send messages that contain encoded processes which can be decoded and invoked when they are received.
The implementation of this feature and the changes to the categorical semantics were written by Norouzbeygi in \cite{No25}. 
The current implementation of CaMPL is available at \url{https://campl-ucalgary.github.io/}.

In this article, we illustrate CaMPL's concurrent type system and the above features through a series of code examples.
We have also summarized these features with the corresponding channel types and process commands in Table~\ref{Table:summary} in Appendix~\ref{sec:summarytable}.

We use \tc{this style} whenever we introduce CaMPL terminology.

\section{A brief overview of CaMPL}
\label{Sec:overview}

As is tradition, we will introduce CaMPL with a ``Hello World!'' example program, shown in Example \ref{ex:helloworld}.
We demonstrate a \tc{process} named \lstinline$helloworld$ printing \lstinline$"Hello World!"$ by sending the string as a message on a \tc{channel} named \lstinline$console$.
The \lstinline$console$ channel has type \lstinline$Console$ which is a special channel type built-in to the compiler to connect any CaMPL program to the terminal from which it is run.
\begin{lstlisting}[label=ex:helloworld, caption=Hello World!, name=helloworld] 
proc helloworld :: | Console => =
	| console => -> do
		hput ConsolePut on console
		put "Hello World!" on console    -- sends message to the console
		hput ConsoleClose on console
		halt console                     -- closes console channel and halts
\end{lstlisting}
The \lstinline$helloworld$ process can be invoked by calling it in the main process, \lstinline$run$, and giving it the \lstinline$console$ channel as follows:
\begin{lstlisting}[name=helloworld] 
proc run =
	| console => -> helloworld( | console => )  -- creates helloworld process
\end{lstlisting}

In Example \ref{ex:serverecho}, we will demonstrate communication between two user-defined processes: \lstinline$client$ and \lstinline$server$. 
The \lstinline$client$ will send a string to \lstinline$server$, and \lstinline$server$ will receive it and echo it back.
The main process will create the \lstinline$client$ and \lstinline$server$ processes, and it will use a \tc{process command} called \lstinline$plug$ to connect them by a channel named \lstinline$ch$.

\newpage
\begin{lstlisting}[label=ex:serverecho, caption=Server echos Client's message]
proc run =
	| => -> plug				-- connects processes by shared channel ch
		client( | => ch )	-- creates client process
		server( | ch => )	-- creates server process
\end{lstlisting}
The CaMPL compiler will infer the type of \lstinline$ch$ using the process' definitions. We will discuss channel types in Section~\ref{Sec:channeltypes}.

The client process has an \tc{output polarity channel} \lstinline$ch$ which connects it to the server.
We discuss polarity in Section~\ref{Sec:polarities}.
Client sends its message with \lstinline$put$, and it receives the server's echo with \lstinline$get$.
Then, it uses \lstinline$halt$ to close the channel with the server and halt.
We can define \lstinline$client$ as follows:
\begin{lstlisting}
proc client =
	| => ch ->
		on ch do
			put "Hello Server!"	-- sends message to the server
			get echo						-- receives server's echo
			halt
\end{lstlisting}

The server process has an \tc{input polarity channel} \lstinline$ch$ which connects it to the client.
It listens to the client and receives the client's message with \lstinline$get$, echos the message back with \lstinline$put$, and finally closes the channel and halts with \lstinline$halt$.
We can define \lstinline$server$ as follows:
\begin{lstlisting}
proc server =
	| ch => ->
		on ch do
			get msg					-- receives client's message
			put msg					-- sends message back
			halt
\end{lstlisting}

These are simple programs that exemplify how CaMPL programs are written.
In fact, the lines of the code in the second program are out of order as the \lstinline$run$ process should be the final process defined in a program.
Furthermore, the runnable program obtained by reordering the lines still would not produce any observable effects since none of the processes are connected to the outside world.
A reader should be left with open questions such as ``what was the type of channel \lstinline$ch$?'' and ``what else can this language do?''
We hope that these questions motivate the reader to continue on to the more complicated examples in the following sections.

\section{Processes}
\label{Sec:Processes}

Processes are the main actors of a CaMPL program. 
A process is specified by the keyword \lstinline$proc$ followed by a process name, an optional type signature, and lists of its variables and channels.

In Example \ref{ex:helloworld}, the type signature of process \lstinline$helloworld$ was ``\lstinline$ :: | Console => $'' which indicates that it has access to one channel of type \lstinline$Console$.
This channel is given the name \lstinline$console$ as indicated by ``\lstinline$ = | console => $'' on the next line.
We will discuss type signatures in Section~\ref{Sec:channeltypes}.

A CaMPL program must have a process with the name \lstinline$run$ in which the execution of the program begins. This process is the main process and the only process which can create \tc{service channels}, such as the \lstinline$Console$, for communication with the outside world.

Consider the process \lstinline$broadcast$ shown in Example \ref{ex:broadcastmsg}.
\begin{lstlisting}[label=ex:broadcastmsg, caption=Process that broadcasts a message from a single source] 
proc broadcast =
	confirmation_msg | source, dest1 => dest2 -> do
		get msg on source								-- receive message from source
		put msg on dest1								-- send message to other processes
		put msg on dest2
		put confirmation_msg on source	-- send confirmation message to source
		close source									  -- close all channels and halt
		close dest1
		halt dest2  -- the last channel is closed with the halt command
\end{lstlisting}
Observe that line 2 consists of three comma-seperated lists: 1)~variables, e.g. \lstinline$confirmation_msg$, 2)~input polarity channels, e.g. \lstinline$source,  dest1$, and 3)~output polarity channels, e.g. \lstinline$dest2$.
\tc{Variables} are instances of sequential types and \tc{channels} are instances of concurrent types.
The channels left of \lstinline$=>$ are \tc{input polarity} channels.
The channels right of \lstinline$=>$ are \tc{output polarity} channels.

Lines 3~-~9 constitute the \tc{process body}.
The channels listed in line 2 are ``in scope'' of the process body which means that the process can perform operations, called \tc{process commands}, on these channels.
The process body may be a single process command, as shown in the \lstinline$run$ processes in Examples~\ref{ex:helloworld} and~\ref{ex:serverecho}, or a \lstinline$do$ block of process commands, as shown in Example \ref{ex:broadcastmsg}.
Process commands performed on the same channel can be grouped together with ``\lstinline$on ch do$'' as in Example~\ref{ex:serverecho} on line 3 of \lstinline$client$.
This allows one to omit a repetitive ``\lstinline$on ch$'' after each command.

\section{Channels}
\label{Sec:Channels}

A process \tc{uses} process commands on a channel to interact with the process plugged into the other end. 
Two processes are connected by (at most) one typed channel that they use with opposite polarities. 
The \tc{type} of a channel defines the interaction that the processes will have over it, and its \tc{polarities} define each process's role in the interaction. 
Thus, the process commands that a process can use on a channel depend on both type and polarity.


\subsection{Channel polarities}
\label{Sec:polarities}

Recall from Section \ref{Sec:Processes} that processes are defined with two lists of the channels in its scope:
input polarity channels and output polarity channels.
A channel's polarity refers to which \tc{end} of the channel the process uses. 

In Example~\ref{ex:serverecho}, the \lstinline$client$ and the \lstinline$server$ are connected by a channel \lstinline$ch$. The \lstinline$client$ uses \lstinline$ch$ with output polarity (the left end) and \lstinline$server$ uses \lstinline$ch$ with input polarity (the right end).
In the first step of the interaction, a message travels from the left to the right.
The \lstinline$client$ uses \lstinline$put$ on its output polarity channel \lstinline$ch$ to send a message. Complementary to \lstinline$put$, \lstinline$server$ uses \lstinline$get$ on its input polarity channel \lstinline$ch$ to receive the message.
In the second step, the message travels in the opposite direction, so the type of \lstinline$ch$ is inferred differently compared to the previous step.

Channel polarities allow one to define unambiguous interactions between processes. 
Without polarities to distinguish their roles, both \lstinline$client$ and \lstinline$server$ could use a \lstinline$get$ command at the same time.
Then, they would both wait for a message to be sent by the other, thereby causing a deadlock. 
This scenario is illustrated in Appendix \ref{Sec: channel polarity deadlock prevention}.

\subsection{Built-in channel types}
\label{Sec:channeltypes}

Built-in channel types define the fundamental interactions that are permitted to occur between processes.
The type of a channel can either be explicitly defined in the type signature of a process, or the compiler can infer the type of a channel based on how the processes on either end use it.

\subsubsection{\lstinline$Put$ and \lstinline$Get$ types}
\label{Sec:putandget}

As mentioned in Section~\ref{Sec:Processes}, process definitions may include a type signature with three comma-separated lists of sequential types, input concurrent types, and output concurrent types.
Variables and channels are bound to types in the order they appear.

We modify \lstinline$server$ from Example~\ref{ex:serverecho} and include its type signature.
To effectively illustrate the types, we redefine \lstinline$server$ to have a \lstinline$server_id$ variable that it will send back to  \lstinline$client$:
\begin{lstlisting}[label=ex:serverid-run, caption=Server replies to Client with server id]
proc server :: Int | Put([Char]|Get(Int|TopBot)) => =		-- type signature
	server_id | ch => ->				-- new server_id variable
		on ch do
			get msg									-- receives msg from client
			put server_id						-- sends server_id back
			halt
\end{lstlisting}
Note that \lstinline$ch$ has type \lstinline$Put([Char]|  Get(Int| TopBot))$ which is constructed inductively. 
The outer-most layer \lstinline$Put([Char]| ...)$ defines the first step of the interaction in which a string \lstinline$msg$ travels from output polarity to input polarity (\lstinline$client$ to \lstinline$server$). Accordingly, in line 4, \lstinline$server$ uses \lstinline$get$ to receive \lstinline$msg$. 
In the next layer \lstinline$Get(Int| ...)$, an integer \lstinline$server_id$ travels from input polarity to output polarity (\lstinline$server$ to \lstinline$client$).
In line 5, \lstinline$server$ uses \lstinline$put$ to send \lstinline$server_id$.
The inner-most layer \lstinline$TopBot$ is the base case and final step. 
We can \lstinline$close$ the channel or, as above, \lstinline$halt$ by closing the final open channel.
Appendix \ref{Sec:ex-put-get} has a complete version of Example \ref{ex:serverid-run}.

\subsubsection{Tensor and Par types}
\label{Sec:tensorandpar}

We will now consider compound channel types that allow bundling two (or more) channels into one.
The types \lstinline$(*)$, called \tc{tensor}, and \lstinline$(+)$, called \tc{par}, come from multiplicative linear logic and allow changes to be made to the network of processes while ensuring no cycles are introduced.
By unbundling the channels, two new channels are created and passed into two new processes.

We modify Example~\ref{ex:serverecho} to demonstrate \lstinline$(*)$.
We define a new process \lstinline$two_clients$ that uses \lstinline$fork$ to unbundle its output channel \lstinline$two_ch$ and create two instances of \lstinline$client$ that both send a message to a modified \lstinline$server$ process. 
This \lstinline$server$ uses \lstinline$split$ to unbundle its input channel \lstinline$two_ch$ into two channels, \lstinline$ch1$ and \lstinline$ch2$. After unbundling, \lstinline$two_ch$ is no longer in scope in either process. 

\newpage
\begin{lstlisting} [label=ex:clientfork, caption=Client forks to create two new processes]
proc client :: | => Put([Char]|TopBot) =
	| => ch ->
		on ch do
			put "Hello Server!"					-- client sends string
			halt
proc two_clients :: | => Put([Char]|TopBot) (*) Put([Char]|TopBot) =
	| => two_ch ->
		fork two_ch as								-- unbundles channels ch1 and ch2
			ch1 -> client( | => ch1)		-- creates two new client processes
			ch2 -> client( | => ch2)		-- that each get one channel
proc server :: | Put([Char]|TopBot) (*) Put([Char]|TopBot) => =
	| two_ch => -> do
		split two_ch into ch1, ch2		-- unbundles channels ch1 and ch2
		on ch1 do											-- interacts with first client
			get msg
			close
		on ch2 do											-- interacts with second client
			get msg
			halt
proc run :: | => =
	| => -> plug
		two_clients( | => two_ch )		-- creates a single two_clients process
		server( | two_ch => )
\end{lstlisting}
A channel of type \lstinline$(*)$ indicates that the process on the left will be replaced by two new processes that are both connected to the process on the right.
What if we wanted the process on the right to be replaced by two new processes instead?
Then, we would use  \lstinline$(+)$ which is dual to \lstinline$(*)$.  For a schematic on the usage of \lstinline$(*)$ and \lstinline$(+)$ with \lstinline$split$ and \lstinline$fork$, see Figures \ref{Fig: Tensor} and \ref{Fig: Par} in Appendix \ref{Sec: Schematic tensor par}.

\subsection{Custom channel types}
\label{Sec:protocols}

We have covered some basic channel types, but our processes cannot send variable numbers of messages.
Consider process \lstinline$forward$ in Example~\ref{ex:forward} that only forwards a single message from \lstinline$source$:
\begin{lstlisting}[label=ex:forward, caption=Process that forwards a message]
proc forward :: | Put([Char]|TopBot) => Put([Char]|TopBot)  =
	 | source => dest -> do
		get msg on source								-- receive message from source
		put msg on dest									-- forward message to dest
		close source										-- close all channels and halt
		halt dest
\end{lstlisting}

To allow \lstinline$forward$ to call itself recursively until all messages are sent, the channel type needs to be able to change from \lstinline$Put$ to \lstinline$TopBot$.
This can be achieved by defining custom channel types, called \tc{protocols} and \tc{coprotocols}, which can be recursive.
Consider the protocol called \lstinline$PassMessages$:

\newpage
\begin{lstlisting} %[label=ex:protocolpassmsgs, caption=Protocol for passing an arbitrary number of messages]

protocol PassMessages(A| ) => S =			-- passes messages of type A
    SendMsg :: Put(A|S) => S						-- handle to send another message
    CloseCh :: TopBot => S							-- handle to finish sending messages
\end{lstlisting}
The above protocol allows an arbitrary number of type \lstinline$A$ messages to be sent on a \lstinline$Put$ channel.
The \tc{handles} represent the set of valid interactions, or session types, that may take place on the channel. 
The handle \lstinline$SendMsg$ sets the channel type to \lstinline$Put$ to allow another message to be sent.
Once all messages are sent, \lstinline$CloseCh$ can be used to change the channel type to \lstinline$TopBot$.
We modify \lstinline$forward$ to use the \lstinline$PassMessages$ protocol to send an arbitrary number of type \lstinline$[Char]$ messages: 
\begin{lstlisting} [label=ex:forwardmsgs, caption=Process that forwards an arbitrary number of messages]
proc forward :: | PassMessages([Char]| ) => PassMessages([Char]| ) =
	| source => dest ->
		hcase source of									-- check whether there is another message
			SendMsg -> do									-- indicates another message will be sent
				get msg on source
				on dest do									-- forward handle and msg
					hput SendMsg		
					put msg
				forward( | source, dest => )-- recurse
			CloseCh -> do									-- indicates all messages have been sent
				close source
				on dest do									-- forward handle and halt
					hput CloseCh	
					halt
 \end{lstlisting}
The sequential type variable \lstinline$A$ is instantiated with \lstinline$[Char]$ on line 1 to send an arbitrary number of strings.
A handle is received on the input polarity channel \lstinline$source$ using \lstinline$hcase$ on line 3.
Process bodies for each handle are defined on lines 4 - 9 and 10 - 14.
On lines 7 and 13, a communication session is initiated on the output polarity channel \lstinline$dest$ by \tc{activating} it with a handle using \lstinline$hput$.
This sets the channel type as specified by the handle. 

Coprotocols are functionally the same as protocols. 
The only difference between protocols and coprotocols is the direction in which the handles travel on the channel, see Figure \ref{Fig: hput and hcase} in Appendix~\ref{Sec:protocol and coprotocol}.
For protocols, handles travel in the same direction as messages on a \lstinline$Put$ channel (\lstinline$hput$ on output and  \lstinline$hcase$ on input).
For coprotocols, handles travel in the same direction as a \lstinline$Get$ channel (\lstinline$hput$ on input and  \lstinline$hcase$ on output). 
For a modified  \lstinline$forward$ process which uses a protocol to receive from \lstinline$src$ and a coprotocol to send on an input polarity \lstinline$dest$, see Appendix \ref{Sec:protocol and coprotocol}.

\section{Controlled non-determinism}
\label{Sec:races}
Recall Example~\ref{ex:clientfork} in which two \lstinline$client$ processes each sent a message to \lstinline$server$. 
Notice that \lstinline$server$ received the messages in a pre-determined order.

Consider the case when \lstinline$server$ echos these messages back.
With the features we have used so far, \lstinline$server$ must have the interactions in a pre-determined order.
It would wait for the first \lstinline$client$ before starting its interaction with the second \lstinline$client$, so the second \lstinline$client$ would also wait for the first \lstinline$client$. This is parallel but not concurrent!
For true concurrency, \lstinline$server$ should dynamically opt to have interactions in the order it receives messages. 
We call this \tc{controlled non-determinism} which uses the \lstinline$race$ process command.
A non-deterministic \lstinline$server$ is shown in Example~\ref{ex:nondetserver}:
\begin{lstlisting} [label=ex:nondetserver, caption=Server echos each client in the order it receives messages]
proc server_deterministic =
	| winner, loser => -> do
		on winner do		-- interacts with client it received a message from first
			get msg
			put msg
			close
		on loser do			-- interacts with other client
			get msg
			put msg
			halt
proc server =
	| two_ch => -> do
		split two_ch into ch1, ch2
		race					-- controlled non-determinism via a race
			ch1 -> server_deterministic( | ch1, ch2 => )		-- client on ch1 wins
			ch2 -> server_deterministic( | ch2, ch1 => )		-- client on ch2 wins
\end{lstlisting}
Races can be held for any number of channels on which a process is waiting for a message.
The \lstinline$race$ on line 14 consists of channels \lstinline$ch1$ and \lstinline$ch2$ as indicated by lines 15 and 16, respectively.
A process body is defined for each channel in the race, and the process will execute according to the winning channel's corresponding process body.
If \lstinline$ch1$ wins, line 15 will execute and \lstinline$server_deterministic$ will be called with \lstinline$ch1$ in the argument for the  \lstinline$winner$ channel.
If \lstinline$ch2$ wins, line 16 will execute and \lstinline$server_deterministic$ will be called with \lstinline$ch2$ in the argument for the \lstinline$winner$ channel.

\section{Higher-order message passing}
\label{sec:homsgs}

In sequential functional programming, higher-order functions take other functions as input or produce functions as output. 
Although higher-order functions are not currently supported in CaMPL, a concurrent analogue, called \tc{higher-order processes}, is.
Higher-order processes encode other processes as sequential data -- called a \tc{higher-order message}, pass a higher-order message, and/or decode and invoke an encoded process.

Example \ref{ex:hoprocs} shows higher-order processes passing \lstinline$helloworld$ from Example~\ref{ex:helloworld}:
\begin{lstlisting} [label=ex:hoprocs, caption=Passing \lstinline{helloworld} between higher-order processes]
proc ho_sender :: | => Put( Store(|Console=>) | TopBot) =
	| => ch -> do
		on ch do
			put store(helloworld)		-- encode helloworld and send as a message
			halt
			
proc ho_receiver :: | Put( Store(|Console=>) | TopBot), Console => =
	| ch, console => -> do
		on ch do
			get stored_process			-- receive message with encoded helloworld
			close
		on console do
			hput ConsolePut
			put ("Server says: Running the stored process")
		use(stored_process)( | console => )		-- decode and invoke helloworld
\end{lstlisting}
Observe that \lstinline$ho_sender$ and \lstinline$ho_receiver$ are connected along a \lstinline$Put(Store(|Console=>) | TopBot)$ type channel.
The built-in sequential type \lstinline$Store$ indicates that a higher-order message is being passed.
The type signature of the encoded process, in this case \lstinline$helloworld$, is indicated within the \lstinline$Store$ type.

To encode a process, the built-in \lstinline$store$ function is used.
This function takes either the name of a previously defined process, as in the above example on line 4, or it takes an anonymous process as an in-line process definition.

To decode and invoke an encoded process, the \lstinline$use$ process command is used as in the above example on line 15.
To \lstinline$use$ an encoded process, a higher-order process must provide instances of variables and channels of the correct types and polarities as specified by the \lstinline$Store$ type signature.

\section{Discussion and future work}

The CaMPL project is still very much in progress.
As such, CaMPL is a proof-of-concept langauge.
On the implementation side, we are actively working on a paper to comprehensively document every programming feature.
We are working on updating the compiler to be able to race protocol/coprotocol handles.
We also want to add features to work with processes that are distributed over multiple devices and connected by a network. We have summarized the programming features discussed in this article, and the corresponding process commands in Table \ref{Table:summary} in Appendix \ref{sec:summarytable}.

On the categorical semantics side, we are working on a precise semantics for non-determinism that considers how races interact with the other features.
We are also considering a categorical semantics for message passing between quantum processes \cite{QMP23}.
This will contribute to the development of programming languages for distributed quantum computing over a quantum internet.

\printbibliography

@mastersthesis{No25,
  author       = {Melika Norouzbeygi},
  title        = {Higher-Order Message Passing in CaMPL},
  school       = {University of Calgary},
  year         = {2025},
  month        = {September},
  type         = {Master's Thesis},
  address      = {Calgary, Alberta, Canada},
  url          = {https://cspages.ucalgary.ca/~robin/Theses/Melika.pdf}
}

@misc{QMP23,
  title        = {Quantum Message Passing Logic (Talk)},
  author       = {Robin Cockett and Priyaa Varshinee Srinivasan},
  year         = {2023},
  howpublished = {\url{https://www.reluctantm.com/gcruttw/fmcs2023/Slides/FMCS_2023_Day_2.pdf}},
  note         = {Accessed: 2026-04-13}
}

@mastersthesis{Ku18,
  author  = {Prashant Kumar},
  title   = {{Implementation of Message Passing Language}},
  school  = {University of Calgary},
  year    = {2018},
  month   = {February},
  type    = {Master's Thesis},
  address = {Calgary, Alberta, Canada},
  url 	  = {https://cspages.ucalgary.ca/~robin/Theses/PrashantKumar.pdf}
}

@mastersthesis{Ye12,
  author  = {Masuka Yeasin},
  title   = {{Linear Functors and their Fixed Points}},
  school  = {University of Calgary},
  year    = {2012},
  address = {Calgary, Alberta, Canada},
  month   = {December},
  type    = {Master's Thesis},
  url 	  = {https://cspages.ucalgary.ca/~robin/Theses/masuka_thesis.pdf}
}

@mastersthesis{Ly18,
  author  = {Reginald Lybbert},
  title   = {{Progress for the Message Passing Logic}},
  school  = {University of Calgary},
  year    = {2018},
  month   = {April},
  type    = {Undergraduate Thesis},
  address = {Calgary, Alberta, Canada},
  url 	  = {https://github.com/campl-ucalgary/campl/blob/main/resources/ProgressForMPL.pdf}
}

@mastersthesis{Po21,
  author  = {Jared Pon},
  title   = {{Implementation Status of CMPL}},
  school  = {University of Calgary},
  year    = {2021},
  month   = {December},
  type    = {Undergraduate Thesis Interim Report},
  address = {Calgary, Alberta, Canada},
  url 	  = {https://github.com/campl-ucalgary/campl/blob/main/resources/502.02A_interim_pon.pdf}
}

@mastersthesis{Po22,
  author  = {Jared Pon},
  title   = {{Redesigning the Abstract Machine for CaMPL}},
  school  = {University of Calgary},
  year    = {2022},
  month   = {April},
  type    = {Undergraduate Thesis},
  address = {Calgary, Alberta, Canada},
  url    = {https://github.com/campl-ucalgary/campl/blob/main/resources/JaredPon_final_report.pdf}
}

@mastersthesis{xanna,
  author  = {Alexanna Little},
  title   = {{Semantics for Non-Determinism in the Categorical Message Passing Language}},
  school  = {University of Calgary},
  year    = {2022},
  month   = {September},
  type    = {PURE Final Assignment: Research Findings and Synthesis},
  address = {Calgary, Alberta, Canada},
  url    = {https://github.com/campl-ucalgary/campl/blob/main/resources/PURE2022_ResearchFindingsSynthesis_Little.pdf}
}

@mastersthesis{xanna2,
  author  = {Alexanna Little},
  title   = {{Formalizing Non-Determinism in the Categorical Message Passing Language}},
  school  = {University of Calgary},
  year    = {2023},
  month   = {April},
  type    = {Undergraduate Thesis},
  address = {Calgary, Alberta, Canada},
  url    = {https://github.com/campl-ucalgary/campl/blob/main/resources/CPSC502F22W23_FinalReport_Little.pdf}
}

@incollection{Howard1980,
  author    = {William A. Howard},
  title     = {The Formulae-as-Types Notion of Construction [Original manuscript from 1969]},
  booktitle = {To H. B. Curry: Essays on Combinatory Logic, Lambda Calculus and Formalism},
  editor    = {Jonathan P. Seldin and J. Roger Hindley},
  publisher = {Academic Press},
  pages     = {479--490},
  year      = {1980},
  isbn      = {978-0-12-349050-6}}

@incollection{Lambek1969,
  author    = {Joachim Lambek},
  title     = {Deductive systems and categories II: Standard constructions and closed categories},
  booktitle = {Category Theory, Homology Theory and their Applications I},
  series    = {Lecture Notes in Mathematics},
  volume    = {86},
  editor    = {P. Hilton},
  publisher = {Springer},
  pages     = {76--122},
  year      = {1969}
}

@incollection{Lambek1972,
  author    = {Joachim Lambek},
  title     = {Deductive systems and categories III: Cartesian closed categories, intuitionist propositional calculus, and combinatory logic},
  booktitle = {Toposes, Algebraic Geometry and Logic},
  series    = {Lecture Notes in Mathematics},
  volume    = {274},
  editor    = {F. W. Lawvere},
  publisher = {Springer},
  pages     = {57--82},
  year      = {1972}
}

@article{CoP07,
  title={The logic of message-passing},
  author={Cockett, Robin and Pastro, Craig},
  journal={Science of Computer Programming},
  volume={74},
  number={8},
  pages={498--533},
  year={2009},
  publisher={Elsevier}
}

@article{CS97,
	author = {Cockett, Robin and Seely, Robert},
	journal = {Journal of Pure and Applied Algebra},
	number = {2},
	pages = {133--173},
	title = {Weakly distributive categories},
	volume = {114},
	year = {1997}}

\newpage 
\appendix

\section{Summary of types and process commands}
\label{sec:summarytable}

The following table summarizes the interactions provided by the built-in channel types, the programming features discussed in this article, and the corresponding process commands.
We denote an argument for a sequential type with \lstinline$A$ and a concurrent channel type with \lstinline$Ch$, \lstinline$Ch1$, and \lstinline$Ch2$.

\begin{table}[h]
\begin{center}
	\quad \begin{tabular}{ |m{3.5cm} | m{7cm} | m{2cm} | m{1.2cm} | }
	\hline
		 Channel type & Description & Process command & Section \\
		 \hline
		 \hline      
		 \lstinline$TopBot$ & Interaction on channel is over.  & \lstinline$close $or \lstinline$halt$ & \ref{Sec:putandget} \\
		 \hline      
		 \lstinline$Neg(Ch)$ & Dual interaction of \lstinline$Ch$ by negating and identifying with another channel.  & \lstinline$neg$ and \lstinline$|=|$  &  \\
		 \hline      
		 \lstinline$Put(A|Ch)$ & Message of type \lstinline$A$ travels from left to right (output to input polarity).  & \lstinline$put$/\lstinline$get$ & \ref{Sec:putandget} \\
		 \hline      
		 \lstinline$Get(A|Ch)$ & Message of type \lstinline$A$ travels from right to left (input to output polarity). & \lstinline$get$/\lstinline$put$ & \ref{Sec:putandget} \\
		 \hline      
		 \lstinline$Ch1(*) Ch2$ & Left process becomes two processes both connected to the right process.  & \lstinline$fork$/\lstinline$split$ & \ref{Sec:tensorandpar} \\
		 \hline
		 \lstinline$Ch1(+) Ch2$ & Right process becomes two processes both connected to the left process.  & \lstinline$split$/\lstinline$fork$ & \ref{Sec:tensorandpar} \\
		 \hline
		 Custom \lstinline$protocol$ & Handles travel from left to right (output to input polarity). & \lstinline$hput$/\lstinline$hcase$ & \ref{Sec:protocols} \\
		 \hline
		 Custom \lstinline$coprotocol$ & Handles travel from right to left (input to output polarity). & \lstinline$hcase$/\lstinline$hput$ & \ref{Sec:protocols} \\
		 \hline
		  N/A & Connect pairs of processes by channels.  & \lstinline$plug$ & \ref{Sec:overview} \\
		 \hline
		  \lstinline$Put(A|Ch)$ or \newline \lstinline$Get(A|Ch)$ & Next process command block selected with controlled non-determinism.  & \lstinline$race$ & \ref{Sec:races} \\
		 \hline
		  N/A & Encode another process in a sequential \lstinline$Store$ type. & \lstinline$store$ &  \ref{sec:homsgs}\\
		 \hline
		  N/A & Decode and invoke the process encoded in a \lstinline$Store$ type. & \lstinline$use$ & \ref{sec:homsgs} \\
		 \hline
	\end{tabular}
	\caption{Summary of channel types and process commands.}
	\label{Table:summary}
\end{center}
\end{table}

\section{Channel polarity example}
\label{Sec: channel polarity deadlock prevention}
Defining the way a process uses a channel by both the type and polarity prevents ambiguity that can cause deadlocks.
For example, suppose we do not consider polarity and rewrite our code as follows:

\newpage
\begin{lstlisting}[label=ex:ambiguityce, caption=Code without channel polarity (this code will not compile)]
 proc run =
	| => -> plug				-- connects processes by shared channel ch
		client( | ch )		-- client process has channel ch (no defined polarity)
		server( | ch )		-- server process has channel ch (no defined polarity)
\end{lstlisting}

We know that we want \lstinline$ch$ to have a type indicating that a message should be sent and received on it.
However, if both processes are defined to first receive a message and then reply, they will both become stuck waiting for a message to be sent by the other:
\begin{lstlisting}
proc client =
	| ch ->
		on ch do
			get msg							-- receives server's message
			put "Hello Server!"	-- sends message to the server
			halt

proc server =
	| ch ->
		on ch do
			get msg							-- receives client's message
			put "Hello Client!"	-- sends message to client
			halt
\end{lstlisting}
This is precisely what a deadlock is.
In this simple example, it is easy to see the mistake, but in more complicated programs, the type system with polarities has the ability to catch these sorts of errors.
Each channel type specifies the complementary roles of the processes on each end of the channel to ensure it is unambiguous.

\section{\lstinline$Put$ and \lstinline$Get$ types example}
\label{Sec:ex-put-get}

\begin{lstlisting}
proc server :: Int | Put([Char]|Get(Int|TopBot)) => =		-- type signature 
	server_id | ch => ->		-- new server_id variable
		on ch do
			get msg
			put server_id				-- sends server_id back
			halt
proc client :: | => Put([Char]|Get(Int|TopBot)) =				-- type signature
	| => ch ->
		on ch do 
			put "Hello Server!"	
			get int					
			halt
proc run =
	| => -> plug						-- connects processes by shared channel ch
		client( | => ch )			-- creates client process
		server( 5 | ch => )		-- creates server process
\end{lstlisting}

\section{Schematic of fork and split}
\label{Sec: Schematic tensor par}

The diagrams in Figures~\ref{Fig: Tensor} and \ref{Fig: Par} visualize the change in the process network on channel types \lstinline$(*)$ and \lstinline$(+)$. We use \lstinline$+$ to denote the output polarity end and \lstinline$-$ to denote the input polarity.
\begin{figure}[h]
\[ \xymatrixcolsep{40mm}
\xymatrix{\begin{tikzpicture}[scale=1.5]
	\begin{pgfonlayer}{nodelayer}
		\node [style=head, thick] (0) at (-3.75, 2.25) {};
		\node [style=none] (1) at (-3.75, 1.25) {};
		\node [style=none] (2) at (-4, 1) {};
		\node [style=none] (3) at (-3.5, 1) {};
		\node [style=none] (4) at (-4, 1.7) {};
		\node [style=none] (5) at (-3.5, 1.7) {};
		\node [style=head, thick] (6) at (0.25, 2.25) {};
		\node [style=none] (7) at (0.25, 1.25) {};
		\node [style=none] (8) at (0, 1) {};
		\node [style=none] (9) at (0.5, 1) {};
		\node [style=none] (10) at (0, 1.7) {};
		\node [style=none] (11) at (0.5, 1.7) {};
		\node [style=none] (12) at (-3, 1.5) {};
		\node [style=none] (13) at (-0.5, 1.5) {};
		\node [style=none] (14) at (-1.75, 1.75) {\lstinline$(*)$};
		\node [style=none] (15) at (-3, 1.25) {+};
		\node [style=none] (16) at (-0.5, 1.25) {-};
		\node [style=none] (17) at (-3.75, 0.25) {Process $A$};
		\node [style=none] (18) at (0.25, 0.25) {Process $B$};
	\end{pgfonlayer}
	\begin{pgfonlayer}{edgelayer}
		\draw [thick] (0) to (1.center);
		\draw [thick] (1.center) to (2.center);
		\draw [thick] (1.center) to (3.center);
		\draw [thick] (4.center) to (5.center);
		\draw [thick] (6) to (7.center);
		\draw [thick] (7.center) to (8.center);
		\draw [thick] (7.center) to (9.center);
		\draw [thick] (10.center) to (11.center);
		\draw (12.center) to (13.center);
	\end{pgfonlayer}
\end{tikzpicture}
 \ar[r]^{\text{ \lstinline$fork$ / \lstinline$split$}~~~~~~}
& 
\begin{tikzpicture}[scale=1.5]
	\begin{pgfonlayer}{nodelayer}
		\node [style=head, thick] (15) at (5.5, 5.25) {};
		\node [style=none] (16) at (5.5, 4.25) {};
		\node [style=none] (17) at (5.75, 4) {};
		\node [style=none] (18) at (5.25, 4) {};
		\node [style=none] (19) at (5.75, 4.7) {};
		\node [style=none] (20) at (5.25, 4.7) {};
		\node [style=head, thick] (21) at (1.5, 6.25) {};
		\node [style=none] (22) at (1.5, 5.25) {};
		\node [style=none] (23) at (1.75, 5) {};
		\node [style=none] (24) at (1.25, 5) {};
		\node [style=none] (25) at (1.75, 5.7) {};
		\node [style=none] (26) at (1.25, 5.7) {};
		\node [style=none] (27) at (4.75, 4.75) {};
		\node [style=none] (28) at (2.25, 5.5) {};
		\node [style=head, thick] (29) at (1.5, 4.5) {};
		\node [style=none] (30) at (1.5, 3.5) {};
		\node [style=none] (31) at (1.75, 3.25) {};
		\node [style=none] (32) at (1.25, 3.25) {};
		\node [style=none] (33) at (1.75, 3.95) {};
		\node [style=none] (34) at (1.25, 3.95) {};
		\node [style=none] (35) at (2.25, 4) {};
		\node [style=none] (36) at (4.75, 4.5) {};
		\node [style=none] (44) at (7, 4.5) {Process $B$};
		\node [style=none] (45) at (0, 5.5) {Process $A1$};
		\node [style=none] (46) at (0, 3.75) {Process $A2$};
		\node [style=none] (47) at (2.25, 5.75) {+};
		\node [style=none] (48) at (2.25, 3.75) {+};
		\node [style=none] (49) at (4.75, 5) {-};
		\node [style=none] (50) at (4.75, 4.25) {-};
	\end{pgfonlayer}
	\begin{pgfonlayer}{edgelayer}
		\draw [thick] (15) to (16.center);
		\draw [thick] (16.center) to (17.center);
		\draw [thick] (16.center) to (18.center);
		\draw [thick] (19.center) to (20.center);
		\draw [thick] (21) to (22.center);
		\draw [thick] (22.center) to (23.center);
		\draw [thick] (22.center) to (24.center);
		\draw [thick] (25.center) to (26.center);
		\draw [in=0, out=180] (27.center) to (28.center);
		\draw [thick] (29) to (30.center);
		\draw [thick] (30.center) to (31.center);
		\draw [thick] (30.center) to (32.center);
		\draw [thick] (33.center) to (34.center);
		\draw [in=0, out=180] (36.center) to (35.center);
	\end{pgfonlayer}
\end{tikzpicture} }
\]
 \caption{Change in process network on \lstinline$(*)$ type}
 \label{Fig: Tensor}
\end{figure}

\begin{figure}[h]
\[ \xymatrixcolsep{40mm}
\xymatrix{\begin{tikzpicture}[scale=1.5]
	\begin{pgfonlayer}{nodelayer}
		\node [style=head, thick] (0) at (-3.75, 2.25) {};
		\node [style=none] (1) at (-3.75, 1.25) {};
		\node [style=none] (2) at (-4, 1) {};
		\node [style=none] (3) at (-3.5, 1) {};
		\node [style=none] (4) at (-4, 1.7) {};
		\node [style=none] (5) at (-3.5, 1.7) {};
		\node [style=head, thick] (6) at (0.25, 2.25) {};
		\node [style=none] (7) at (0.25, 1.25) {};
		\node [style=none] (8) at (0, 1) {};
		\node [style=none] (9) at (0.5, 1) {};
		\node [style=none] (10) at (0, 1.7) {};
		\node [style=none] (11) at (0.5, 1.7) {};
		\node [style=none] (12) at (-3, 1.5) {};
		\node [style=none] (13) at (-0.5, 1.5) {};
		\node [style=none] (14) at (-1.75, 1.75) {\lstinline$(+)$};
		\node [style=none] (15) at (-3, 1.25) {+};
		\node [style=none] (16) at (-0.5, 1.25) {-};
		\node [style=none] (17) at (-3.75, 0.25) {Process $A$};
		\node [style=none] (18) at (0.25, 0.25) {Process $B$};
	\end{pgfonlayer}
	\begin{pgfonlayer}{edgelayer}
		\draw [thick] (0) to (1.center);
		\draw [thick] (1.center) to (2.center);
		\draw [thick] (1.center) to (3.center);
		\draw [thick] (4.center) to (5.center);
		\draw [thick] (6) to (7.center);
		\draw [thick] (7.center) to (8.center);
		\draw [thick] (7.center) to (9.center);
		\draw [thick] (10.center) to (11.center);
		\draw (12.center) to (13.center);
	\end{pgfonlayer}
\end{tikzpicture}
 \ar[r]^{\text{ \lstinline$split$ / \lstinline$fork$}~~~~~~}
& 
\begin{tikzpicture}[scale=1.5]
	\begin{pgfonlayer}{nodelayer}
		\node [style=head, thick] (15) at (1.5, 5.25) {};
		\node [style=none] (16) at (1.5, 4.25) {};
		\node [style=none] (17) at (1.25, 4) {};
		\node [style=none] (18) at (1.75, 4) {};
		\node [style=none] (19) at (1.25, 4.7) {};
		\node [style=none] (20) at (1.75, 4.7) {};
		\node [style=head, thick] (21) at (5.5, 6.25) {};
		\node [style=none] (22) at (5.5, 5.25) {};
		\node [style=none] (23) at (5.25, 5) {};
		\node [style=none] (24) at (5.75, 5) {};
		\node [style=none] (25) at (5.25, 5.7) {};
		\node [style=none] (26) at (5.75, 5.7) {};
		\node [style=none] (27) at (2.25, 4.75) {};
		\node [style=none] (28) at (4.75, 5.5) {};
		\node [style=head, thick] (29) at (5.5, 4.5) {};
		\node [style=none] (30) at (5.5, 3.5) {};
		\node [style=none] (31) at (5.25, 3.25) {};
		\node [style=none] (32) at (5.75, 3.25) {};
		\node [style=none] (33) at (5.25, 3.95) {};
		\node [style=none] (34) at (5.75, 3.95) {};
		\node [style=none] (35) at (4.75, 4) {};
		\node [style=none] (36) at (2.25, 4.5) {};
		\node [style=none] (44) at (0, 4.5) {Process $A$};
		\node [style=none] (45) at (7, 5.5) {Process $B1$};
		\node [style=none] (46) at (7, 3.75) {Process $B2$};
		\node [style=none] (47) at (2.25, 5) {+};
		\node [style=none] (48) at (2.25, 4.25) {+};
		\node [style=none] (49) at (4.75, 5.75) {-};
		\node [style=none] (50) at (4.75, 3.75) {-};
	\end{pgfonlayer}
	\begin{pgfonlayer}{edgelayer}
		\draw [thick] (15) to (16.center);
		\draw [thick] (16.center) to (17.center);
		\draw [thick] (16.center) to (18.center);
		\draw [thick] (19.center) to (20.center);
		\draw [thick] (21) to (22.center);
		\draw [thick] (22.center) to (23.center);
		\draw [thick] (22.center) to (24.center);
		\draw [thick] (25.center) to (26.center);
		\draw [in=-180, out=0] (27.center) to (28.center);
		\draw [thick] (29) to (30.center);
		\draw [thick] (30.center) to (31.center);
		\draw [thick] (30.center) to (32.center);
		\draw [thick] (33.center) to (34.center);
		\draw [in=180, out=0] (36.center) to (35.center);
	\end{pgfonlayer}
\end{tikzpicture} }
\]
 \caption{Change in process network on \lstinline$(+)$ type}
 \label{Fig: Par}
\end{figure}

\section{Protocol and coprotocol example}
\label{Sec:protocol and coprotocol}

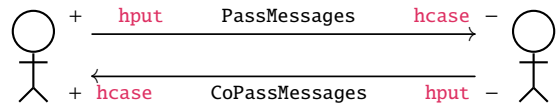
\begin{figure}[h]
\[  \xymatrix{  
 \blackman  \quad
  \ar@<8pt>[rrrrr]^{+~~~~~\ttc{hput} \qquad {\tt PassMessages} \qquad \ttc{hcase}~~- }
 \ar@<-8pt>@{<-}[rrrrr]_{+~~\ttc{hcase} \qquad {\tt CoPassMessages} \qquad \ttc{hput}~~-} 
  &  & & & & \quad \blackman   }  
 \]
 \caption{Direction of handles on protocols and coprotocols}
 \label{Fig: hput and hcase}
\end{figure}

Protocol handles travel in the same direction as messages on a \lstinline$Put$ channel, so \lstinline$forward$ is receiving handles on \lstinline$source$.
Coprotocol handles travel in the same direction as messages on a \lstinline$Get$ channel, so \lstinline$forward$ is sending handles on \lstinline$dest$.

To demonstrate a usage of coprotocols, we consider  process \lstinline$forward$ which forwards a single message from \lstinline$source$ over an input channel \lstinline$dest$:
\begin{lstlisting}[label=ex:coforward, caption=Process that forwards a message]
proc forward :: | Put([Char]|TopBot), Get([Char]|TopBot) =>   =
	 | source, dest => -> do
		get msg on source								-- receive message from source
		put msg on dest									-- forward message to dest
		close source										-- close all channels and halt
		halt dest
\end{lstlisting}

We need \lstinline$forward$ to send an arbitrary number of strings on a \lstinline$Get$ channel. Furthermore, we want handles to travel in the same direction as the messages, so we define a coprotocol version of \lstinline$PassMessages$ (in Section \ref{Sec:protocols} 
) called \lstinline$CoPassMessages$ to forward the messages on \lstinline$dest$.

\begin{lstlisting}
coprotocol S => CoPassMessages (A| ) =
    CoSendMsg ::  S => Get(A|S)
    CoCloseCh :: S => TopBot
\end{lstlisting}
Notice that the \tc{state variable} \lstinline$S$ is on the left side of \lstinline$=>$ in the \lstinline$CoPassMessages$ definition instead of the right side in \lstinline$PassMessages$.
We use different handle names than in \lstinline$PassMessages$ because all handle names must be globally unique.

We modify \lstinline$forward$ to use \lstinline$PassMessages$ and \lstinline$CoPassMessages$ as follows:
\begin{lstlisting} [label=ex:coforwardmsgs, caption=Process that forwards an arbitrary number of messages]
proc forward :: | PassMessages([Char]| ), CoPassMessages([Char]| ) => =
	| source, dest => ->
		hcase source of				-- check whether there is another message
			SendMsg -> do				-- protocol handle that indicates another message
				get msg on source
				on dest do
					hput CoSendMsg	-- send coprotocol handle to forward msg
					put msg
				forward( | source, dest => ) -- recurse
			CloseCh -> do				-- protocol handle that indicates all messages sent
				close source
				on dest do
					hput CoCloseCh	-- send handle to indicate all messages sent
					halt
 \end{lstlisting}

\end{document}